\documentclass[superscriptaddress,aps,journal=prx,preprint]{revtex4-2}
\usepackage{graphicx}
\usepackage{epstopdf}
\usepackage{bm,amsmath,amssymb} 
\usepackage{wasysym}
\usepackage{gensymb}
\usepackage{color, soul} 
\usepackage{dcolumn}   
\usepackage{multirow}
\usepackage[colorlinks=true, linkcolor=black, citecolor=black, urlcolor=blue]{hyperref}

\newcommand{\note}[1]{{\color{black}{{#1}}}}

\newcommand{\Unuct}{$\Delta U_{\rm nuc}^*$}
\newcommand{\OP}{O$^{p}$}
\newcommand{\ONP}{O$^{np}$}
\newcommand{\Ec}{$\mathcal{E}_c$}

\usepackage{lineno}
\begin{document}
\title{Theoretical lower limit of coercive field in ferroelectric hafnia}
\author{Jiyuan Yang}
\thanks{These authors contributed equally}
\affiliation{Department of Physics, School of Science, Westlake University, Hangzhou, Zhejiang 310024, China}
\author{Jing Wu}
\thanks{These authors contributed equally}
\affiliation{Department of Physics, School of Science, Westlake University, Hangzhou, Zhejiang 310024, China}
\author{Jingxuan Li}
\thanks{These authors contributed equally}
\affiliation{State Key Laboratory of Advanced Welding and Joining of Materials and Structures, School of Materials Science and Engineering, Harbin Institute of Technology, Shenzhen, 518055, China}
\author{Chao Zhou}
\affiliation{State Key Laboratory of Advanced Welding and Joining of Materials and Structures, School of Materials Science and Engineering, Harbin Institute of Technology, Shenzhen, 518055, China}
\author{Yang Sun}
\affiliation{Department of Physics, Xiamen University, Xiamen 361005, China}
\author{Zuhuang Chen}
\email{zuhuang@hit.edu.cn}
\affiliation{State Key Laboratory of Advanced Welding and Joining of Materials and Structures, School of Materials Science and Engineering, Harbin Institute of Technology, Shenzhen, 518055, China}
\author{Shi Liu}
\email{liushi@westlake.edu.cn}
\affiliation{Department of Physics, School of Science, Westlake University, Hangzhou, Zhejiang 310024, China}
\affiliation{Institute of Natural Sciences, Westlake Institute for Advanced Study, Hangzhou, Zhejiang 310024, China}

\begin{abstract}
The high coercive field ($\mathcal{E}_c$) of hafnia-based ferroelectrics presents a major obstacle to their applications. The ferroelectric switching mechanisms in hafnia that dictate $\mathcal{E}_c$, especially those related to domain nucleation in the Nucleation-Limited-Switching (NLS) model and domain wall motion in the Kolmogorov-Avrami-Ishibas (KAI) model, have remained elusive. We develop a deep-learning-assisted multiscale approach, incorporating atomistic insights into the critical nucleus, to predict both NLS- and KAI-type coercive fields.
The theoretical NLS-type \Ec~values agree with previous experimental results as well as our own measurements and also exhibit the correct thickness scaling 
for films between 3 and 20 nm. Combined theoretical and experimental investigations reveal that the giant \Ec~in hafnia-based ferroelectrics arises from the ultra-thin geometry, which confines switching to the NLS mechanism. We predict that the theoretical lower limit for KAI-type \Ec~is 0.1 MV/cm arsing from mobile domain walls. The activation of KAI-type switching to achieve lower \Ec~ is supported by our experimental demonstration of a low coercive field of 1 MV/cm in a 60 nm ferroelectric (HfO$_2$)$_n$/(ZrO$_2$)$_n$ ($n=3$ unit cells) superlattices. These findings establish a comprehensive framework for understanding ferroelectric switching in hafnia and highlight the potential of geometry and domain-wall engineering to achieve low-\Ec~devices.
\end{abstract}


\maketitle
\newpage
\date{\today}

\section{Introduction}
Ferroelectric hafnia is emerging as a viable candidate for incorporating ferroelectric capabilities into semiconductor technologies for its proven compatibility with silicon~\cite{Boscke11p102903, Schroeder22p653} and scale-free ferroelectricity~\cite{Lee20p1343, Cheema20p478}. The discovery of switchable spontaneous polarization in this fluorite-structured binary oxide has also significantly advanced our understanding of ferroelectric physics, an area historically focused on perovskite oxides.  For example, recent investigations have uncovered several unique aspects of hafnia-based ferroelectrics, including its metastable nature compared to the bulk monoclinic phase~\cite{Materlik15p134109}, the presence of flat phonon bands~\cite{Lee20p1343}, and intertwined polar-nonpolar structural polymorphism with oxygen vacancy diffusion~\cite{Nukala21p630,Ma23p096801}.

However, the widespread application of ferroelectric hafnia-based devices is mainly hindered by reliability issues, especially their limited endurance~\cite{Silva23p089201}. Unlike devices based on perovskite ferroelectrics like Pb(Zr,Ti)O$_3$ and BiFeO$_3$, which can be made close to fatigue-free~\cite{Araujo95p627}, hafnia-based devices, depending on their architectures, typically exhibit poor endurance, with switching cycles ranging between $10^{5}$ and $10^{12}$~\cite{Park20p240904,Pesic16p4601, Yurchuk14p3699,Ambriz17p13262}. This falls short of the required $>10^{15}$ cycles for embedded random access memory. Such a detrimental issue is attributed to the high coercive field (\Ec, the electric field needed to switch the polarization) of hafnia-based ferroelectrics (typically $>$1 MV/cm in polycrystalline thin films and $>$2 MV/cm in epitaxial thin films)~\cite{Mimura18p102901,Wei18p1095, Song21p12224}. 
Repeated applications of high electric fields near the breakdown strength to reverse the polarization are expected to accelerate breakdown, causing rapid performance deterioration. 
Reducing \Ec~without sacrificing spontaneous polarization has therefore become a pressing challenge for hafnia-based ferroelectrics.

The $Pca2_1$ orthorhombic phase ($o$-phase) of HfO$_2$ is widely acknowledged as the ferroelectric phase in thin films~\cite{ Huan14p064111,Sang15p162905,Materlik15p134109, Park15p1811}. However, the atomic-level switching mechanisms affecting \Ec~in this phase remain elusive, due to the complexity of competing switching pathways and the diverse types of domain walls, which are interfaces separating differently polarized domains.
Previous studies using zero-Kelvin density functional theory (DFT) have examined polarization switching at the unit-cell level~\cite{Huan14p064111,Ma23p256801} and the behavior of specific types of DWs~\cite{Lee20p1343, Ding20p556, Choe21p8,Wu23p226802}, often using the nudged elastic band (NEB) method. This technique requires manually constructing an initial switching pathway, introducing potential bias as the chosen pathway may influence results. 
\note{Critically, the switching barriers computed in these studies effectively correspond to a spinodal decomposition-like mechanism~\cite{Ricinschi98p477,Ducharme00p175}, where all dipoles switch homogeneously and simultaneously throughout the crystal. Such a mechanism is unrealistic for macroscopic samples, where the probability of uniform switching is negligible due to the vast number of dipoles. Instead, it is well-established that ferroelectric switching occurs through a thermally activated nucleation-and-growth process~\cite{Shin07p881,Liu16p360}, where small regions (nuclei) of reversed polarization form and grow, eventually propagating throughout the material. Capturing this nucleation-driven mechanism is essential for accurate predictions of \Ec.}

\note{
Ferroelectric switching via the nucleation mechanism is commonly modeled using the Kolmogorov-Avrami-Ishibashi (KAI) model~\cite{Ishibashi71p506} and the Nucleation Limited Switching (NLS) model~\cite{Tagantsev02p214109}. The KAI model describes switching dynamics with two key parameters: the constant nucleation rate $R$ which defines the probability of forming new three-dimensional (3D) reverse domains in unswitched regions, and the domain wall velocity $v$, which characterizes the speed of domain wall propagation. In this model, switching occurs through a combination of domain nucleation and domain wall propagation. Microscopically, domain wall propagation involves the nucleation of two-dimensional (2D) reverse domains on the wall, followed by their growth, as proposed by Miller and Weinreich~\cite{Miller59p1176}. Since 3D nucleation is energetically more difficult than 2D nucleation~\cite{Landauer57p227}, the rate-limiting step in the KAI model is the 2D nucleation-and-growth process on the domain wall~\cite{Shin07p881,Liu16p360}.

In contrast, the NLS model emphasizes the nucleation of reverse domains, accounting for spatial variations in the nucleation rate $R$. It assumes a site-dependent nucleation rate, leading to a broad distribution of nucleation probabilities. Switching in the NLS model occurs through independent, stochastic, and and spatially inhomogeneous nucleation of reverse domains at different sites, with the rate-limiting step being the 3D nucleation-and-growth process, which is distinct from the 2D domain-wall-driven process in the KAI model.

The applicability of these models depends on the material system and geometry. The KAI model is suited for bulk ferroelectrics, where domain wall motion dominates, while the NLS model better describes ferroelectric thin films, where limited volume and high surface-to-bulk ratio enhance the role of domain nucleation.
Experimental studies support the NLS model for hafnia-based thin films, where switching is nucleation-driven~\cite{Materano20p262904,Zhou24p2893}.
However, both the KAI and NLS models are empirical frameworks primarily used to fit the time dependence of switching currents. While they offer valuable insights into macroscopic switching behavior, they do not directly reveal atomic-level switching mechanisms. Moreover, a theoretical approach to predict \Ec~by integrating first-principles insights into the KAI or NLS frameworks is currently unavailable.

Here, we establish a multiscale approach that bridges atomistic simulations with the NLS and KAI switching models, enabling accurate predictions of \Ec~values in ferroelectric hafnia using only first-principles parameters. By leveraging a deep neural network-based force field trained exclusively on first-principles data for hafnia~\cite{Zhang18p143001, Wu21p024108}, we determine the critical nucleus with molecular dynamics (MD) for 3D nucleation in the NLS model and 2D nucleation in the KAI model.
We further develop a universal Landau–Ginzburg–Devonshire (LGD) analytical model, applicable to both 3D and 2D nucleation, to predict \Ec~values for NLS- and KAI-type switching. Remarkably, the predicted thickness-dependent NLS-type \Ec~values show excellent agreement with prior experimental data and our own measurements for epitaxial hafnia thin films grown by pulsed laser deposition (PLD). Our combined experimental-theoretical analysis reveals that the giant \Ec~in hafnia thin films arises from the ultra-thin geometry, which restricts switching to the NLS mechanism.
Moreover, our model predicts that enabling KAI-type switching in hafnia, via mobile domain walls, would reduce \Ec~to levels comparable with Pb(Ti,Zr)O$_3$, with a theoretical lower limit of 0.1 MV/cm in thin films. This is corroborated by our experimental results from PLD-grown ferroelectric HfO$_2$/ZrO$_2$ superlattices of 60~nm thickness, achieving a low coercive field of $\approx$1.0 MV/cm. This study provides valuable insights into the origin of \Ec~in hafnia-based ferroelectrics and outlines strategies for optimization through geometry and domain-wall engineering. 
}

\section{Results and Discussion}
\subsection{Order parameter and reference phase}
\begin{figure}
	\begin{center}
		\includegraphics[width=0.9\textwidth]{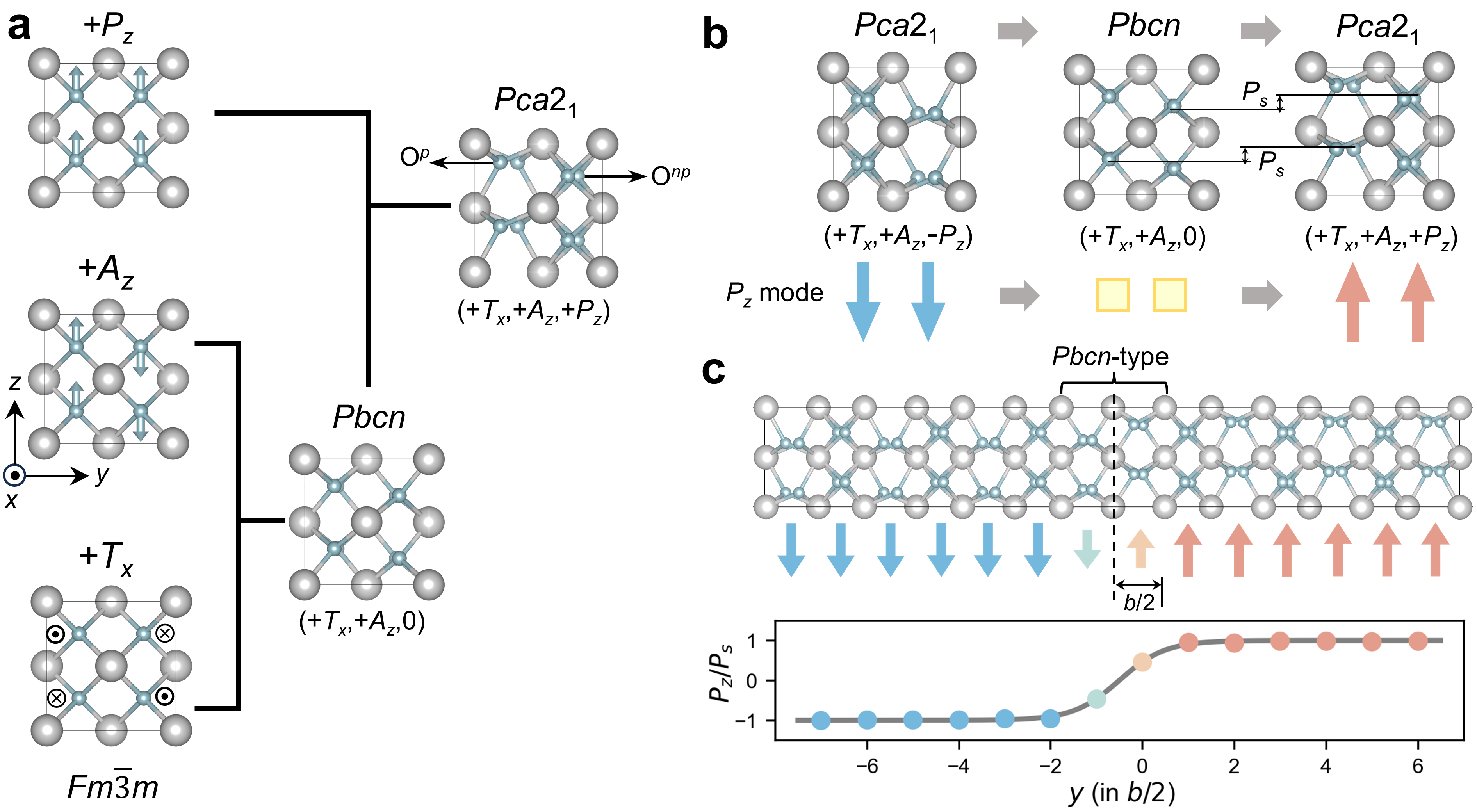}
	\end{center}
	\caption{\textbf{Symmetry mode analysis and local order parameter.} \textbf{a,} HfO$_2$ phases connected through three lattice modes, $T_x$, $A_z$, and $P_z$, in the cubic phase. \textbf{b,} Polarization reversal in the $Pca2_1$ unit cell driven by an downward electric field $\mathcal{E}$. Both \OP~and \ONP~atoms are displaced by $P_s$.
    \textbf{c,} Schematic of crystal structure for a $Pbcn$-type 180$\degree$ domain wall and the profile of mode $P_z$ at each oxygen site across the wall. 
 } 
 \label{Mode}
\end{figure}

The unit cell of $Pca2_1$ HfO$_2$ consists of alternating nonpolar layers of fourfold-coordinated oxygen ions (\ONP) and polar layers of threefold-coordinated oxygen ions (\OP) (Fig.~\ref{Mode}a).
This phase is considered as a multi-order parameter system, as indicated by previous studies~\cite{Delodovici21p064405, Zhou22peadd5953, Aramberri23p95}, where three primary modes, tetragonal ($T_x$), antipolar ($A_z$), and polar ($P_z$), are involved for the transition from the cubic $Fm\bar{3}m$ phase to the $Pca2_1$ phase with polarization aligned along the $z$-axis. In the cubic reference phase, the $T_x$ mode is characterized by antiparallel $x$-displacements of neighboring oxygen atoms, the $A_z$ mode involves antiparallel $z$-displacements, and the $P_z$ mode is defined by parallel $z$-displacements of all oxygen atoms. 
Starting from the cubic phase, the successive distortions following the $T_x$, $A_x$, and $P_x$ modes lead to tetragonal ($t$) $P4_2/nmc$, antipolar $Pbcn$, and polar $Pca2_1$ phases, respectively. 
The unit-cell configuration of $Pca2_1$ can thus be described by a mode vector $(T_x, A_z, P_z)$.

The alternating \OP~and \ONP~atoms complicates the analysis of polarization switching mechanism, particular within the framework of Landau phase transition. We find that the mode of $P_z$ relative to the $Pbcn$ phase is the most convenient order parameter. Using the $Pbcn$ phase as the reference, rather than the cubic or tetragonal phase, is justified for the following reasons. First, the minimum energy path (MEP) for polarization reversal, identified with DFT-based NEB calculations, reveals that the pathway with the antipolar $Pbcn$ phase as the transition state has a lower barrier (0.25 eV) compared to the pathway involving the $t$-phase as the transition state~\cite{Zhou22peadd5953}. Second, 
the $Pca2_1$ phase is connected to the $Pbcn$ phase $(T_x, A_z, 0)$ through a single soft mode $P_z$, as pointed out by Raeliarijaona and Cohen~\cite{Raeliarijaona23p094109}.
Third, as shown in Fig.~\ref{Mode}b, 
both \ONP~and \OP~atoms are displaced equally from their reference sites in the $Pbcn$ phase, allowing them to be treated equivalently in terms of the $P_z$ mode. Similarly for a 180$\degree$domain wall depicted in Fig.~\ref{Mode}c, the $P_z$ profile  closely resembles the local polarization profile of a typical 180$\degree$ domain wall in perovskite ferroelectrics. Finally, our MD simulations (discussed below) demonstrate that domain nucleation and domain wall motion are primarily driven by changes in the $P_z$ mode, while $T_x$ and $A_z$ remain nearly constant. Therefore, choosing the $P_z$ mode relative to the $Pbcn$ phase as the order parameter greatly simplifies the analysis by eliminating the complexity arising from the alternating \ONP~and \OP~layers.

\subsection{Domain nucleation from MD simulations}

The unit-cell switching barrier of 0.25 eV computed with DFT-based NEB corresponds to a spinodal decomposition-like switching process, where all dipoles flip simultaneously across the crystal. Such a process is highly unfeasible, as it yields a value of \Ec~exceeding 10 MV/cm~\cite{Chen23p11,Maeda17p1,Qi20p214108}, far greater than the experimentally observed range of 2–5 MV/cm~\cite{Wei18p1095,Lyu19p220,Zhou24p2893,Song21p12224,Song20p3221,Lyu19p6224}.
To address this discrepancy, we develop a multiscale approach to accurately predict the value of \Ec~for NLS-type switching in single-crystal single-domain $Pca2_1$ HfO$_2$.  Our approach begins with finite-field MD simulations to identify the critical nucleus at representative electric field strengths. The geometrical features of the nucleus are subsequently incorporated into a Landau-Ginzburg-Devonshire (LGD) phenomenological framework, allowing for efficient estimation of critical nucleus size ($N^*$), the size at which the nucleus has an equal probability of growing or shrinking, under arbitrary field conditions. The value of NLS-type \Ec~is defined as the minimum strength required to make the critical nucleus size comparable to the film thickness, thereby establishing a direct thickness-dependent relationship for \Ec. 

\begin{figure}
	\begin{center}
		\includegraphics[width=0.6\textwidth]{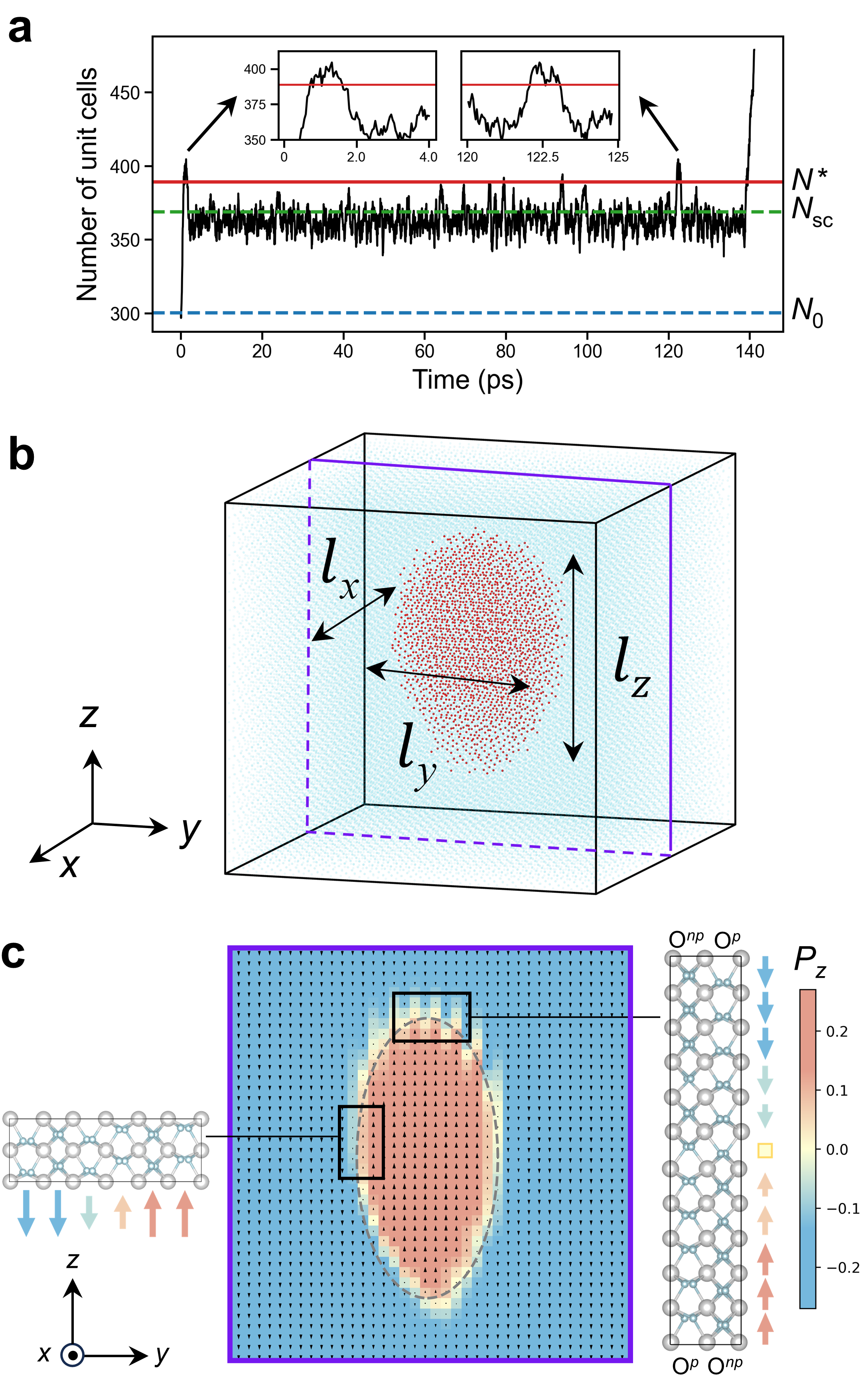}
	\end{center}
	\caption{\textbf{Domain nucleation from molecular dynamics.} \textbf{a,} Nucleus size ($N$) as a function of time during the nucleation process, obtained using the MD-based $r$PEM method at 300~K and $\mathcal{E}=2.5$~MV/cm. The blue and green dashed lines indicate the size of the persistent 2D embryo ($N_0$) and the sub-critical nucleus ($N_{\rm sc}$), where the harmonic bias is removed, respectively. The critical size ($N^*$) is marked with a red solid line. Insets highlight plateaus where $N$ fluctuates around $N^*$. \textbf{b,} Visualization of the critical nucleus, showing a 3D ellipsoidal geometry. Only oxygen atoms within the nucleus are displayed for clarity. \textbf{c,} $P_z$ profile of the nucleus in the $yz$ plane, with lattice structures depicted for the top and side interfaces.  
 } 
 \label{3Dnuc}
\end{figure}

Determining $N^*$ at low electric fields using conventional MD simulations is challenging due to the inherently low nucleation probability. We employ a revised persistent-embryo method ($r$PEM, see Fig.~S2)~\cite{Sun18p085703} that allows for efficient and unbiased determination of $N^*$. 
Specifically, a 3D embryo with $N_0$ switched oxygen atoms is initially embedded in the unswitched single domain, and a tunable harmonic potential is applied to each oxygen atoms in the embryo to prevent back-switching. 
The potential is gradually weakened and completely removed when the nucleus size ($N$) surpasses a subcritical threshold ($N_{\rm sc}$, where $N_0<N_{\rm sc}<N^*$). As shown in Fig.~\ref{3Dnuc}a, the plateau in the $N(t)$ curve then corresponds to the period over which the nucleus size fluctuates around $N^*$.

At 300 K and under an applied electric field ($\mathcal{E}$) of 2.5 MV/cm along the $z$ axis, the MD-based $r$PEM method reveals a critical nucleus with a 3D ellipsoidal geometry (Fig.~\ref{3Dnuc}b), characterized by lateral dimensions of $l_x^* \approx l_y^* = 3.4$ nm and a vertical dimension of $l_z^* =7.2$ nm. 
Figure~\ref{3Dnuc}c presents the $P_z$ profile of the nucleus in the $yz$ plane, along with lattice structures displayed for the top and side interfaces.  
The interface of the nucleus displays  crystallographic features depending on the orientation. Along the $z$ axis, the unit cells have \OP~atoms gradually evolving to \ONP~atoms (and vice versa), effectively avoiding the formation of high-energy head-to-head or tail-to-tail configurations of \OP~(\ONP) atoms. In contrast, the interfaces in both the $yz$ and $xz$ planes adopt the $Pbcn$-like phase, where antiparallel \OP~atoms create a transitional region between the nucleus and the parent domain.

\subsection{LGD 3D nucleation model}

The geometrical features of the 3D nucleus are used to construct an LGD nucleation model. 
When the $Fm\bar{3}m$ or $P4_2/nmc$ phase is chosen as the reference, the total free energy of the single-domain $Pca2_1$ phase must be expressed as a Taylor series expansion in terms of $P_z$, $T_x$, and $A_z$. In comparison, using the $Pbcn$ phase as the nonpolar reference simplifies the free energy expression to a Taylor series in $P_z$ alone, as the terms involving $T_x$ and $A_z$ are constant during the transition of $Pbcn\rightarrow Pca2_1$.
This is also consistent with results from MD simulations,
which show that nucleation is primarily driven by changes in $P_z$. Consequently, the nucleation energy can be effectively reduced to a function of $P_z$ without compromising predictive accuracy. Moreover, we later find that expressing the nucleation energy in terms of various interfacial energies implicitly incorporates contributions from all order parameters.

In the presence of $\mathcal{E}$, the energy of a nucleus relative to the single-domain state can be expressed as:
\begin{equation}
\Delta U_{\mathrm{nuc}}=\Delta U_{\mathrm{V}}+\Delta U_{\mathrm{I}}
\end{equation}
with 
\begin{equation}
    \Delta U_{\mathrm{V}} = - \mathcal{E} \eta \iiint_V\mathrm{d}V \left (P_z^{\rm nuc}(x,y,z)-P_z^{\rm SD}(x,y,z) \right) 
\end{equation}
and
\begin{equation}
\Delta U_{\mathrm{I}} =  \iiint_V \mathrm{d}V  \left (  U_{\mathrm{loc}}(P_z^{\mathrm{nuc}}) + U_x(P_z^{\mathrm{nuc}}) + U_y(P_z^{\mathrm{nuc}}) + U_z(P_z^{\mathrm{nuc}})  - U_{\mathrm{loc}}(P_z^{\rm SD}) \right )
\end{equation}
Here, $\Delta U_{\mathrm{V}}$ represents the coupling between the electric field and order parameter $P_z$, while $\Delta U_{\mathrm{I}}$ accounts for the energy penalty associated with the creation of new interfaces during nucleation. The parameter $\eta$ is a scaling factor that relates the mode magnitude $P_z$ to the local polarization. The spacial profiles of $P_z$ for a domain with and without a 3D nucleus are represented by $P_z^{\mathrm{nuc}}(x,y,z)$ and $P_z^{\mathrm{SD}}(x,y,z)$, respectively. The term $\Delta U_{\mathrm{I}}$ comprises two contributions: the change in local energy $U_{\rm loc}$ and the gradient energy $U_j$ (where $j = x, y, z$). The change in local energy, $\Delta U_{\rm loc} = \iiint_V\mathrm{d}V[U_{\mathrm{loc}}(P_z^{\mathrm{nuc}}) - U_{\mathrm{loc}}(P_z^{\rm SD})]$, results from unit cells located at the nucleus interface have $P_z$ deviated from the ground-state value ($P_s$). This energy change can be expressed as $\Delta U_{\rm loc} = \iiint_V\mathrm{d}VK_{\rm loc}\left[1 - \left(P_z^{\rm nuc}/P_s\right)^2\right]^2$, where $K_{\rm loc}$ is a LGD coefficient (see Supplementary Sect.~IV). The gradient energy originates from the spatial gradient ($\partial_jP_z^{\rm nuc}$) at the nucleus boundary and is expressed as $U_j = g_j (\partial_jP_z^{\rm nuc})^2$ for $j = x, y, z$, where $g_j$ is the gradient coefficient. 

The $P_z^{\mathrm{nuc}}$ profile for a domain containing a 3D ellipsoidal nucleus, defined by the dimensions $l_x \times l_y \times l_z$ and the diffuseness parameters $\delta_x$, $\delta_y$, and $\delta_z$, can be analytically approximated in spherical coordinates as:
\begin{equation}
P_z^{\mathrm{nuc}}(\rho, \theta, \varphi)=P_s \tanh \left( \frac{\rho-l_\theta/2}{\delta_\theta / 2}\right)
\end{equation}
where $l_\theta=l_xl_z/\sqrt{(l_x\cos\theta)^2+(l_z\sin\theta)^2}$ represents the  radial length of the nucleus, and $\delta_\theta$ is the diffuseness along $\rho$ at the nucleus boundary (see Fig.~\ref{NLSEc}a). The value of $P_z^{\mathrm{SD}}$ is constant, given by $P_z^{\mathrm{SD}}(\rho, \theta, \varphi)=-P_s$. 

Substituting the profiles of $P_z^{\mathrm{nuc}}$ and $P_z^{\mathrm{SD}}$ in to Equation (3) yields an analytical expression for the nucleation energy:
\begin{equation}
\Delta U_{\rm nuc} = -\frac{\pi}{3}l_x^2l_z\eta P_s\mathcal{E}+\int_{0}^{2\pi}\mathrm{d}\varphi\int_{0}^{\pi}\sigma_x\sqrt{\sin^2\theta+\left(\frac{\sigma_z}{\sigma_x}\right)^2\cos^2\theta}~(l_\theta/2)^2\sin\theta\mathrm{d}\theta,
\end{equation}
where $\sigma_j$ is the interfacial energy for a 180$^\circ$ domain wall perpendicular to $j$-axis (see detailed derivations in  Supplementary Sect.~IV). Remarkably, the only material-specific parameters needed to compute $\Delta U_{\rm nuc}$ are $P_s$ and various domain wall energies, which can be readily obtained using DFT and/or model potential. This can be intuitively understood, as both $K_{\rm loc}$ and $g_j$ can be expressed in terms of $\sigma_j$ and $P_s$.
Moreover, since the domain wall energy $\sigma_j$ computed using DFT inherently includes contributions from all order parameters, Equation (5) naturally accounts for these effects, extending beyond a sole dependence on $P_z$. The temperature dependence of the LGD nucleation model is incorporated through $P_s(T)=\sqrt{\frac{T_c-T}{T_c}}P_s$, where $T_c=900$~K (from MD simulations~\cite{Wu23p144102}) and $P_s=0.27~$\AA~at 0~K (from DFT calculations). 

For a given $\mathcal{E}$, the critical nucleus size is determined by identifying the stationary condition of $\Delta U_{\rm nuc}$. As shown in Fig.~\ref{NLSEc}b, the critical dimensions $l_x^*$ and $l_z^*$ predicted by the LGD model across multiple field strengths exhibit excellent agreement with the results obtained from MD-based $r$PEM simulations, validating the accuracy and robustness of the nucleation model.

\subsection{NLS-type coercive fields from first principles}

\begin{figure}
	\begin{center}
		\includegraphics[width=0.6\textwidth]{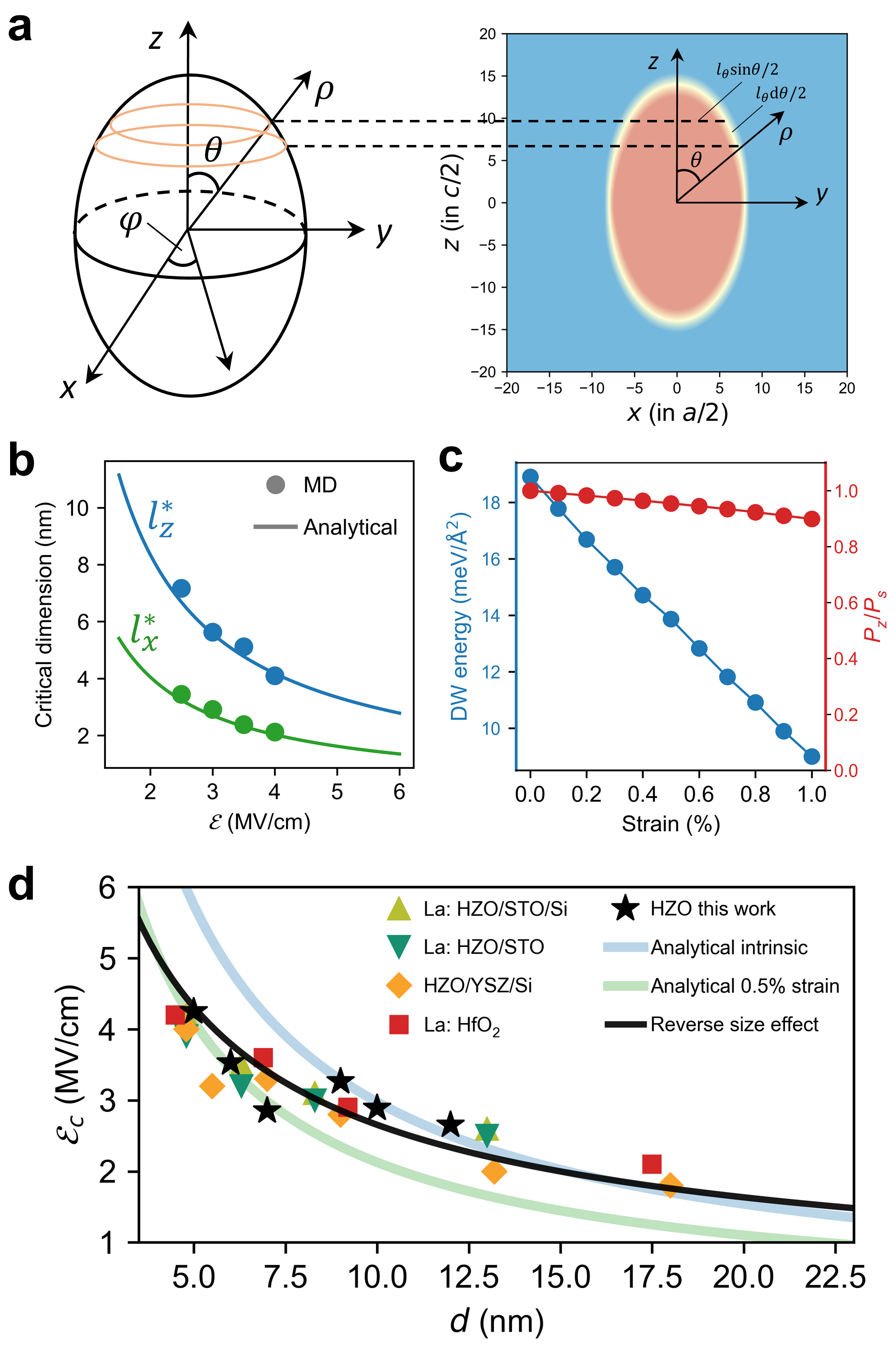}
	\end{center}
	\caption{\textbf{NLS-type coercive fields in hafnia thin films from multiscale simulations and experiments.} \textbf{a,} Schematic of a 3D ellipsoidal nucleus in spherical coordinates. \textbf{b,} Comparison of critical dimensions, $l_x^*$ and $l_z^*$, predicted by the LGD model and MD simulations across various electric fields. \textbf{c,} Effect of tensile strain on the $Pbcn$-type domain wall energy and $P_z$ mode magnitude. \textbf{d,} Theoretical thickness-dependent NLS-type coercive fields compared with experimental results for epitaxial thin films of La-doped HfO$_2$~\cite{Song21p12224} and Hf$_{0.5}$Zr$_{0.5}$O$_2$ (HZO) grown on common substrates such as SrTiO$_3$ (STO)~\cite{Song20p3221} and yttria-stabilized zirconia (YSZ)~\cite{Lyu19p6224}, as well as our own measurements. 
 } 
 \label{NLSEc}
\end{figure}

We now apply the LGD 3D nucleation model to predict the magnitude of intrinsic \Ec~in thin films of HfO$_2$, assuming that ferroelectric switching primarily occurs via nucleation, similar to the NLS model.
Given that experimentally ferroelectric hafnia-based thin films often adopt a [111]-orientation~\cite{Fina21p1530}, we define \Ec~as the minimum electric field along the [111] direction required to make the critical nucleus height, $l_z^*$, comparable to the film thickness, $l_z^* \approx d$. Despite this simple assumption, the intrinsic \Ec~values predicted by our nucleation model (transparent blue line in Fig.~\ref{NLSEc}d) align remarkably well with experimental values reported in the literature~\cite{Song21p12224,Song20p3221,Lyu19p6224}, as well as our own measurements for Hf$_{0.5}$Zr$_{0.5}$O$_2$ (HZO) thin films, particularly when $d \geq 9$~nm. 

For ultrathin films with thicknesses below 7 nm, the theoretical predictions overestimate \Ec~compared to experimental values. This overestimation is resolved by  accounting for the reverse size effect observed in experiments, where the out-of-plane lattice spacing ($d_{111}$) increases with decreasing film thickness~\cite{Cheema20p478}. Specifically, we find that a 0.5\% tensile strain reduces the $Pbcn$-type 180$\degree$ domain wall energy by 16\%, while the value of $P_s$ decreases by only 4\% (Fig.~\ref{NLSEc}c). By incorporating the effect of 0.5~\% tensile strain on $\sigma_j$ into Equation (5), the theoretical \Ec~values (transparent green line in Fig.~\ref{NLSEc}d) closely match experimental results for $d \leq 7$~nm. 
Notably, the theoretical \Ec~is $\approx$4 MV/cm for $d=5$~nm, an excellent agreement with experimental values for La-doped HfO$_2$~\cite{Song21p12224} and Hf$_{0.5}$Zr$_{0.5}$O$_2$ films~\cite{Song20p3221,Lyu19p6224}. Moreover, perfect agreement between theory and experiment (black line in Fig.~\ref{NLSEc}d) for film thickness ranging from 5~nm to 20~nm is achieved by introducing the reverse size effect as a thickness-dependent strain that decreases linearly from 0.5\% at $d = 5$~nm to zero for $d \geq 10$~nm. This assumption is further supported by detailed microstructural analyses of HZO films with varying thicknesses, as elaborated in the following section.

\subsection{Origin of the scaling of coercive field with thickness}

\begin{figure}
	\begin{center}
		\includegraphics[width=0.9\textwidth]{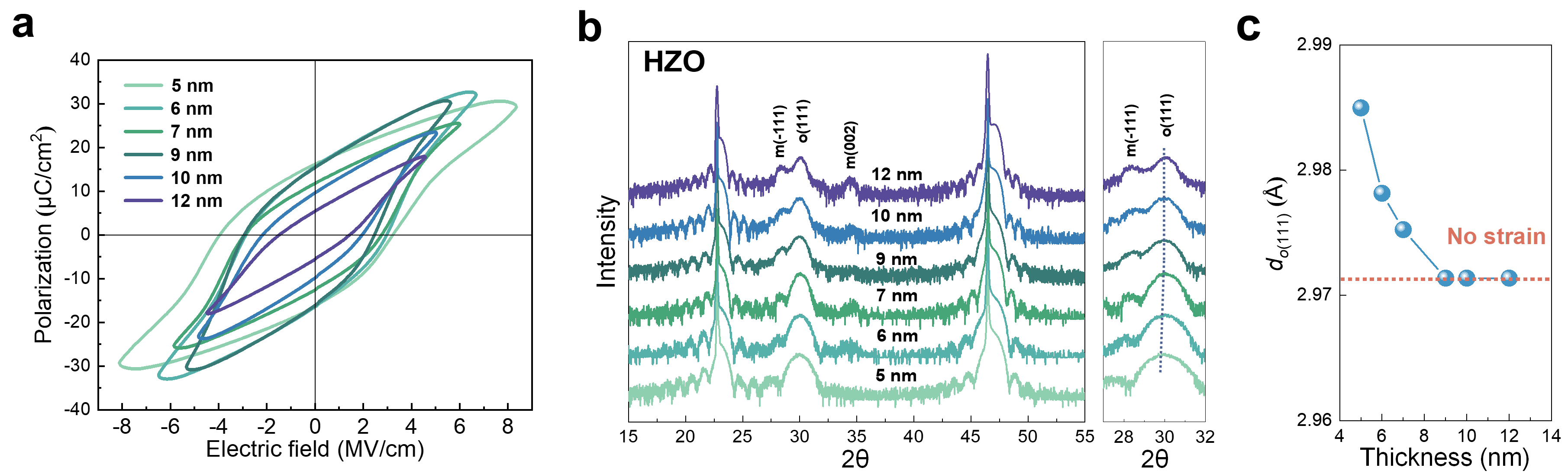}
	\end{center}
	\caption{\textbf{Electric and structural characterizations of PLD-grown HZO films.} \textbf{a,} Polarization-electric field loops  for HZO films of varying thickness. \textbf{b,} XRD $\theta$-2$\theta$ patterns showing the (111)$_o$ reflection of the ferroelectric orthorhombic phase near 30$^\circ$. The (111)$_o$ peak shifts by less than 0.14$^\circ$ with increasing thickness. \textbf{c,} Out-of-plane lattice spacing $d_{111}$ as a function of thickness.
 } 
 \label{Thinfilm}
\end{figure}

In predicting the thickness-dependent NLS-type \Ec, we assume that the polarization of a single domain is unaffected by thickness variations, consistent with the scale-free ferroelectricity of hafnia~\cite{Lee20p1343}. To validate this, we perform structural characterization of HZO films (5-12 nm thick) deposited on (001)-oriented La$_{0.67}$Sr$_{0.33}$MnO$_3$ (LSMO)-buffered SrTiO$_3$ (STO) substrates using PLD. All films exhibit well-defined polarization-electric field hysteresis loops (Fig.~\ref{Thinfilm}a). 
Figure~\ref{Thinfilm}b presents the X-ray diffraction (XRD) $\theta$-2$\theta$ patterns for these films. Along with $(00l)$ reflections from the STO substrates and LSMO electrodes, all films exhibit a peak near 30$^\circ$, corresponding to the (111)$_o$ reflection of the ferroelectric $o$-phase of HfO$_2$. Thicker films also show a peak at 28$^\circ$, attributed to the (11-1)$_m$ plane of the nonpolar monoclinic phase. The (111)$_o$ peak shifts by less than 0.14$^\circ$ with increasing thickness, while the remnant polarization  decreases by over 50\% (Fig.~\ref{Thinfilm}a). These XRD and hysteresis loop measurements collectively demonstrate that the thickness-dependent reduction in macroscopic polarization arises primarily from changes in the volume fraction of the ferroelectric $o$-phase, rather than significant changes in the intrinsic polarization of the ferroelectric domains, supporting our use of scale-free ferroelectricty for \Ec~predictions.   

The thickness dependence of intrinsic theoretical \Ec~arises naturally from the requirement that the critical nucleus size is comparable to the film thickness, leading to a scaling relationship, $\mathcal{E}_c\propto d^{-1}$. Experimental coercive fields in thin films thicker than 9~nm indeed follow this scaling behavior but show substantial deviations in ultrathin films. By introducing a thickness-dependent tensile strain into our nucleation model, we recover the correct scaling for films with thicknesses ranging from 3 to 20~nm. As shown in Fig.~\ref{Thinfilm}c, the (111) lattice spacing decreases with increasing film thickness and saturates at a value of 2.97~\AA~beyond 10~nm. This corresponds to an out-of-plane tensile strain of approximately 0.5\% in 5~nm films, strongly supporting the assumptions underlying our theoretical predictions.

Overall, our nucleation model provides a \textit{minimal yet robust} framework to explain the thickness-dependent \Ec~observed across a wide range of experiments, including polycrystalline thin films grown by atomic layer deposition (ALD)~\cite{Kim14p192903,Park14p072902,Park17p9973} and epitaxial films grown by PLD~\cite{Wei18p1095,Lyu19p220,Zhou24p2893,Song21p12224,Song20p3221,Lyu19p6224}. Furthermore, the strain's thickness dependence, which varies with fabrication methods, accounts for the diverse scaling relationships reported in the literature~\cite{Materano20p262904,Mimura18p102901}.

\subsection{180$^\circ$ domain wall motions from MD simulations}

Our combined theoretical and experimental investigation reveals that the giant \Ec~observed in ferroelectric hafnia-based thin films originates from NLS-type behavior, where 3D nucleus formation constitutes the rate-limiting step. We propose that activating KAI-type switching, governed by domain wall motion, could significantly reduce \Ec. 
To explore this design principle, we first use MD simulations to demonstrate that domains with pre-existing domain walls reverse under much lower fields compared to single domains. Furthermore, the LGD model, informed by MD-based $r$PEM insights into 2D nucleation at the wall, predicts a theoretical lower limit for the  intrinsic KAI-type \Ec~to be as low as 0.1 MV/cm.

Here, we focus on the $Pbcn$-type 180$\degree$ domain wall, which features antiparallel \OP~atoms at the wall, as illustrated in Fig.~\ref{Mode}c. It is important to note that the $Pbcn$-type domain wall differs from the extensively studied $Pbca$-type DW~\cite{Ding20p556,Zhao22p064104,Wu23p226802,Lee20p1343}, where unit cells at the wall adopt the $Pbca$ phase. Specifically, the $Pbcn$-type wall separates $(+T_x, +A_z, +P_z)$ and $(+T_x, +A_z, -P_z)$ domains, while the $Pbca$-type wall separates  $(+T_x, +A_z, +P_z)$ and $(-T_x, -A_z, -P_z)$ domains (see Fig.~S1). Our DFT calculations show that, although the $Pbca$-type wall is the thermodynamically most stable 180$\degree$ wall, it has a prohibitively large motion barrier of 1.3~eV~\cite{Lee20p1343,Choe21p8}, which is far higher than the motion barrier of 0.035~eV for the $Pbcn$-type wall. Our MD simulations confirm that the $Pbca$-type wall is essentially immobile, showing no movement even under under a 7 MV/cm field applied for 1 ns, while the $Pbcn$-type wall is mobile and switches within 1~ns at a low field of 0.36~MV/cm.
\note{We note that, in this context, the field strength primarily serves as a measure of the ease of domain wall motion. However, it is not equivalent to the coercive field, as the coercive field may also depend on the frequency of the applied electric field.}

\begin{figure}
	\begin{center}
		\includegraphics[width=0.9\textwidth]{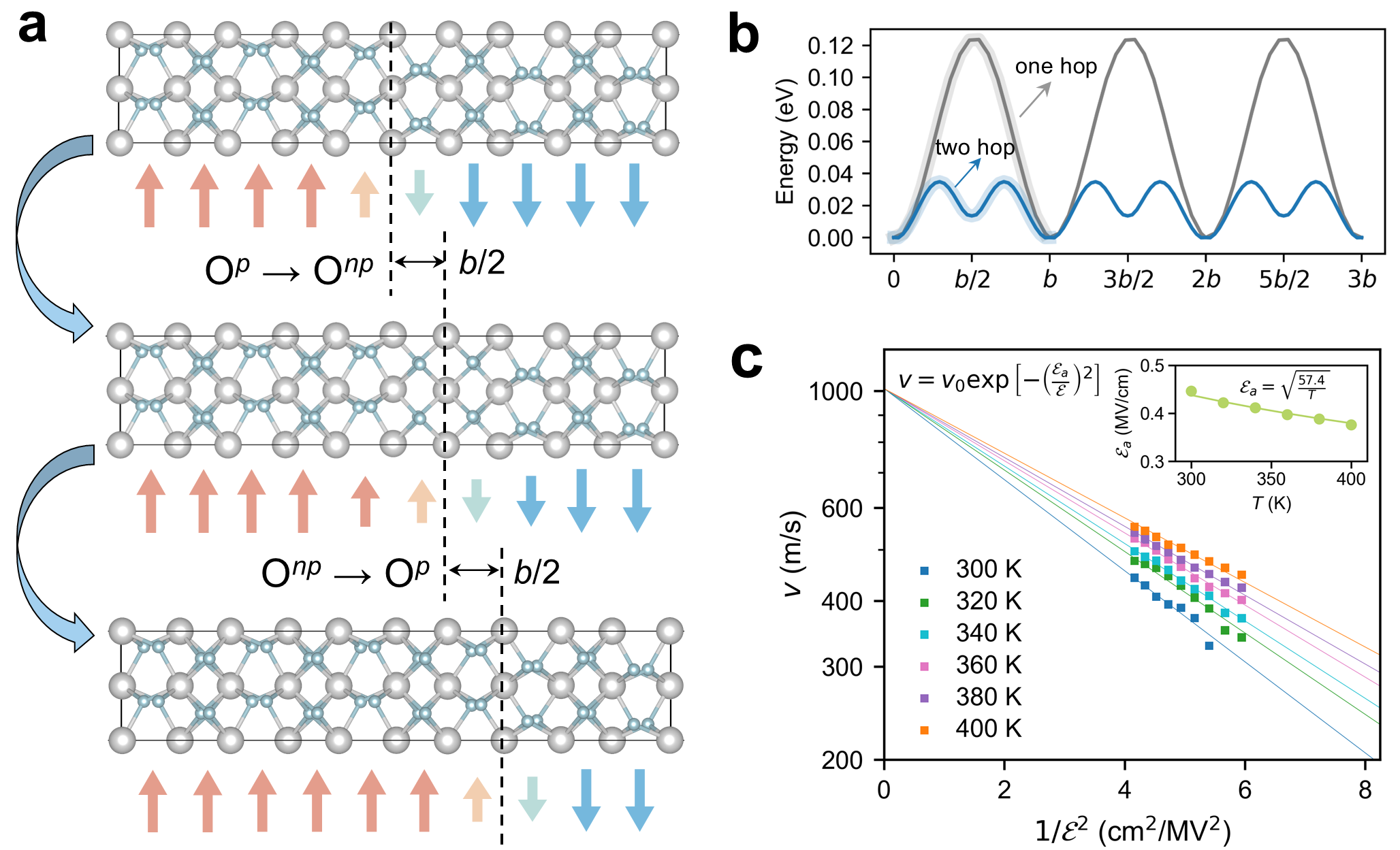}
	\end{center}
	\caption{\textbf{Domain wall motions from MD simulations.} \textbf{a,} Mechanism for the motion of a $Pbcn$-type 180$\degree$ wall deduced from MD simulations. \textbf{b,} Minimum energy pathways from DFT-based NEB calculations for the two-hop mechanism shown in \textbf{a,} and a hypothetical one-hop mechanism. \textbf{c,} Plot of $\ln(v)$ versus $1/\mathcal{E}^2$ at varying temperatures. The inset shows that the temperature dependence of the activation field is well described by $\mathcal{E}_a = \sqrt{T_{\rm C0}/T}$, supporting $\mu=2$.
 } 
 \label{DW}
\end{figure}

The mechanism of domain wall motion extracted from finite-field MD simulations at 300~K is illustrated in Fig.~\ref{DW}a. A distinctive feature is that the lateral one-unit-cell ($b$) translation of the $Pbcn$-type wall occurs through two sequential half-unit-cell ($b/2$) hops. This behavior is markedly different from that observed in perovskite ferroelectrics, where DWs typically advance by an entire unit cell at a time~\cite{Shin07p881, Liu16p360}. 
Specifically, as shown in Fig.~\ref{DW}a, under a downward $\mathcal{E}$, 
the wall shifts by $b/2$ through \OP$\rightarrow$\ONP, followed by another $b/2$ shift via \ONP$\rightarrow$\OP. 
The intermediate state features a $t$-type domain wall where the unit cells at the interface adopts the tetragonal $P4_2/nmc$ phase.
This two-hop mechanism is corroborated by DFT-based NEB calculations. As plotted in Fig.~\ref{DW}b, the MEP displays a characteristic two-hop profile, with a higher energy barrier for the first hop followed by a shallower one for the second. In contrast, the one-hop pathway involving simultaneous \OP$\rightarrow$\ONP~and \OP$\rightarrow$\ONP~results in a much higher energy barrier of 0.125~eV. 


The velocities ($v$) for the $Pbcn$-type wall are quantified using MD simulations across various temperatures ($T$) and field strengths. 
The $\mathcal{E}$-dependence of $v$ can be described with a creep process~\cite{Tybell02p097601, Jo09p045701}:
\begin{equation}
v=v_0\exp\left[-\frac{\mathcal{U}}{k_BT}\left(\frac{\mathcal{E}_{\rm C0}}{\mathcal{E}}\right)^{\mu}\right],
\label{creep}
\end{equation}
where $v_0$ is the domain wall velocity under an infinite field, $k_B$ is Boltzmann's constant, and 
$\mathcal{U}$ and $\mathcal{E}_{\rm C0}$ represent the characteristic energy barrier and electric field at zero Kelvin, respectively; 
$\mu$ is the creep exponent, which depends on the dimensionality of the interface and the universality class of the disorder landscape pinning the interface~\cite{Lemerle98p849}. Equation~(\ref{creep}) can be reformulated as:
\begin{equation}
v=v_0\exp\left[-\left(\frac{\mathcal{E}_a}{\mathcal{E}}\right)^{\mu}\right],
\label{merz}
\end{equation}
where $\mu=1$ corresponds to the famous Merz's law~\cite{Merz54p690,Miller60p1460}, and $\mathcal{E}_a$ is the $T$-dependent activation field. By comparing Equations~(\ref{creep}) and (\ref{merz}), $\mathcal{E}_a$ is given as $\mathcal{E}_a=(T_{\rm C0}/T)^{1/\mu}$ with $T_{\rm C0}=(\mathcal{U}/k_B)\mathcal{E}_{\rm C0}^\mu$. Remarkably, all velocity data fits well with $\mu=2$. This is evidenced by the linear relationship between $\ln v$ and $1/\mathcal{E}^2$ and an accurate  representation of $\mathcal{E}_a$'s  $T$-dependence as $\mathcal{E}_a=\sqrt{T_{\rm C0}/T}$, as shown in Fig.~\ref{DW}c and the inset. A creep exponent of 2 is higher than the well-known value of $\mu=0.25$ for 1D magnetic domain walls in ultrathin magnetic films~\cite{Lemerle98p849} and $\mu=1$ for 2D ferroelectric domain walls in typical perovskite ferroelectrics~\cite{Shin07p881, Liu16p360}. 

\subsection {LGD 2D nucleation model}

\begin{figure}
	\begin{center}
		\includegraphics[width=0.9\textwidth]{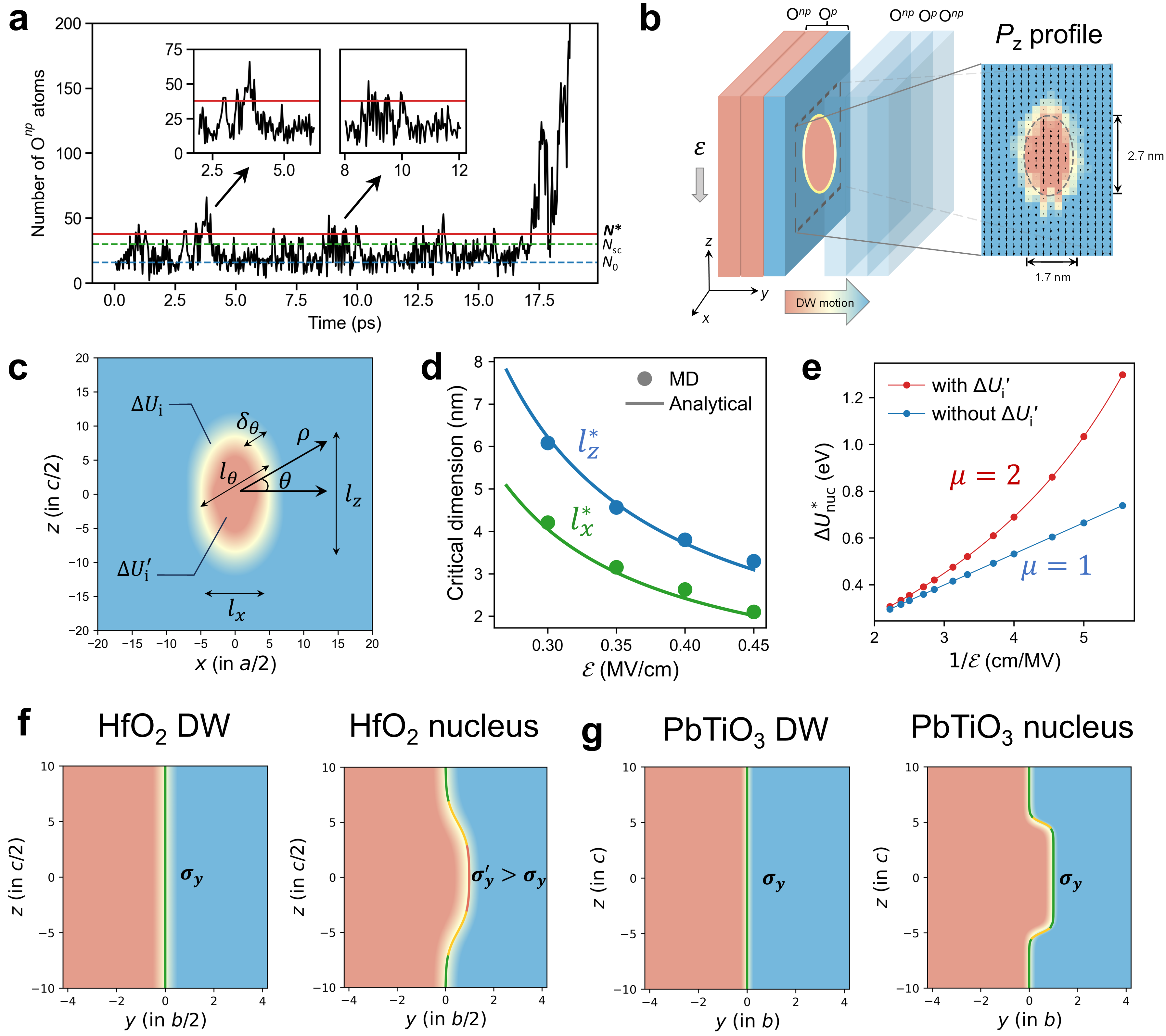}
	\end{center}
	\caption{\textbf{2D nucleation at the domain wall from MD simulations.} \textbf{a,} Nucleus size ($N$) in the number of \ONP~atoms versus time during the nucleation process at a $Pbcn$-type wall, obtained using the MD-based $r$PEM method at 300~K and $\mathcal{E}=0.45$~MV/cm. \textbf{b,} Simulated $P_z$ profiles of a quasi-2D half-unit-cell-thin critical nucleus. The corresponding analytical $P_z$ profile in the $xz$ plane generated using Equation (\ref{Pznuc}) are depicted in \textbf{c}. \textbf{d,} Comparison of critical dimensions, $l_x^*$ and $l_z^*$, predicted by the LGD model and MD simulations across various electric fields. \textbf{e,} Field dependence of critical nucleation barrier $\Delta U_{\rm nuc}^*$ simulated with and without $\Delta U_{\rm i}'$. Comparison of 2D nucleation at a 180$\degree$ wall in \textbf{f,} HfO$_2$ and \textbf{g,} PbTiO$_3$. The nucleation at a $Pbcn$-type wall of HfO$_2$ involves the change of domain wall type to the $t$-type, associated with the change of domain wall energy from $\sigma_y$ to $\sigma_y'$.
 } 
 \label{2Dnuc}
\end{figure}

We now develop a LGD model to explain the origin of  $\mu=2$.
The MD-based $r$PEM method is used to determine the size ($N^*$), measured as number of \ONP~atoms, and the shape of the critical nucleus at the wall (see Methods). Figure~\ref{2Dnuc}a displays the $N(t)$ curve for the $Pbcn$-type wall at 300~K and $\mathcal{E}=0.45$~MV/cm, revealing multiple plateaus, from which the average of $N^*$ is determined to be 38. Consistent with the two-hop mechanism discussed above, the critical nucleus mainly comprises a single layer of \ONP~atoms and exhibits an elliptical shape in the $xz$ plane. When viewed in the $P_z$ profile (Fig.~\ref{2Dnuc}b), its dimensions are $l_x^* = 1.7$~nm (width), $l_z^* = 2.7$~nm (height), and $l_y^* = 0.25$~nm (thickness).

The energy of a quasi-2D nucleus at the wall is derived using the same approach as that outlined for 3D nucleation. 
The nucleation energy is give by 
   $ \Delta U_{\rm nuc} = \Delta U_{\mathrm{V}} + \Delta U_{\mathrm{I}}$
with \begin{equation}
    \Delta U_{\mathrm{V}} = - \mathcal{E} \eta \int_{-\infty}^{\infty} \mathrm{d}z \int_{-\infty}^{\infty} \mathrm{d}y \int_{-\infty}^{\infty} \mathrm{d}x \left[ P_z^{\mathrm{nuc}}(x,y,z) - P_z^{\mathrm{DW}}(x,y,z) \right],
\end{equation}
and
\begin{equation}
\begin{aligned}
\Delta U_{\mathrm{I}} = & \int_{-\infty}^{\infty} \mathrm{d}z \int_{-\infty}^{\infty} \mathrm{d}y \int_{-\infty}^{\infty} \mathrm{d}x \left\{ \left[ U_{\mathrm{loc}}(P_z^{\mathrm{nuc}}) + U_x(P_z^{\mathrm{nuc}}) + U_y(P_z^{\mathrm{nuc}}) + U_z(P_z^{\mathrm{nuc}}) \right] - \right. \\
& \left. \left[ U_{\mathrm{loc}}(P_z^{\mathrm{DW}}) + U_y(P_z^{\mathrm{DW}}) \right] \right\}.
\end{aligned}
\end{equation}
Here, $P_z^{\mathrm{nuc}}$ is the $P_z$ profile for a wall containing a nucleus and $P_z^{\mathrm{DW}}$ is the profile for a nucleus-free wall. For an elliptical nucleus of size $l_x \times l_y \times l_z$, with diffuseness parameters $\delta_x$, $\delta_y$, and $\delta_z$, its boundary within the $xz$ plane is conveniently described with polar coordinates $l_\theta$ and $\theta$, with $l_\theta = l_x l_z/\sqrt{(l_x \sin \theta)^2 + (l_z \cos \theta)^2}$.
The $P_z$ profile for a wall containing a nucleus can then be approximated as 
\begin{equation}
P_z^{\mathrm{nuc}}(y,\rho,\theta) = P_s  g(y,l_y,\delta_y) f(\rho,l_\theta,\delta_\theta) + P_z^{\mathrm{DW}},
\label{Pznuc}
\end{equation}
where $g(y, l_y, \delta_y)$ characterizes the evolution of  $P_z$ along the $y$ axis in the presence of a nucleus, $f(\rho,l_\theta,\delta_\theta)$ is the evolution of  $P_z$ in the $xz$ plane, and $P_z^{\mathrm{DW}}=P_s\tanh\frac{y}{b/2}$. 
Figure~\ref{2Dnuc}c shows the profile of $P_z^{\mathrm{nuc}}(x,y,z)$ in the $xz$ plane generated using the analytical Equation (\ref{Pznuc}), agreeing well with the profile from MD.

The value of $\Delta U_{\rm V}$ is straightforwardly obtained by integrating the difference between $P_z^{\mathrm{nuc}}$ and $P_z^{\mathrm{DW}}$, multiplied by the field strength. For $\Delta U_{\rm I}$, 
we once again derive an expression that depends solely on domain wall energies and $P_s$:
\begin{equation}
\begin{aligned}
\Delta U_\mathrm{I} &=\Delta U_\mathrm{i}' + \Delta U_\mathrm{i} \\
&=  \frac{1}{8}\int_{0}^{2\pi} \sigma_w (l_\theta-\delta_\theta)^2 \mathrm{d}\theta
+ \frac{1}{2} \int_{0}^{2\pi} \sigma_\theta \delta_\theta l_\theta \mathrm{d}\theta 
\end{aligned}
\end{equation}
where $\sigma_w$ represents the change in the interfacial energy 
within the $xz$ plane and $\sigma_\theta$ denotes the $\theta$-dependent interfacial energy for the diffusive interfaces along $\rho$ (Fig.~\ref{2Dnuc}c). Both $\sigma_w$ and $\sigma_\theta$ can be related to conventional domain wall energies (see details in Supplementary Sect.~IV). We emphasize that $\Delta U_\mathrm{i}'$ results from the change in domain wall type during the formation of a half-unit-cell-thin nucleus, a hallmark of the two-hop mechanism where the intermediate state features a $t$-type wall (Fig.~\ref{DW}a). Consequently, $\sigma_w$ is proportional to the domain wall energy difference between the $Pbcn$-type and $t$-type walls. This energy penalty is unique to domain walls in HfO$_2$. In contrast, nucleation at domain walls in perovskite ferroelectrics, such as PbTiO$_3$, does not involve a change in the domain wall type (see Fig.~\ref{2Dnuc}g). 

As demonstrated in Fig.~\ref{2Dnuc}d, the dimensions of the 2D critical nucleus predicted by the LGD 2D nucleation model show remarkable agreement with the results from MD-based $r$PEM. Interestingly, we find that $\Delta U_{\rm nuc}^*$ scales with $1/\mathcal{E}^2$ (Fig.~\ref{2Dnuc}e), indicating the nucleation rate $J\propto \exp(-\frac{\Delta U_{\rm nuc}^*}{k_BT})$ can be reformulated as:
\begin{equation}
    J=J_0\exp\left[-\left(\frac{\mathcal{E}_{a, n}}{\mathcal{E}}\right)^2\right],
\end{equation}
where $J_0$ is a pre-exponential factor and  $\mathcal{E}_{a, n}$ is the effective activation field for nucleation. 
This further corroborates the finding that the creep exponent is $\mu=2$ in Equation~(\ref{merz}). 
A series of model calculations are performed using a nucleation model without $\Delta U_\mathrm{i}'$ to assess the impact of  this unique energy term on the field dependence of $\Delta U_{\rm nuc}^*$. We observe that omitting $\Delta U_\mathrm{i}'$ yields a linear relationship between $\Delta U_{\rm nuc}^*$ and $1/\mathcal{E}$. These results demonstrate  $\Delta U_\mathrm{i}'$ is the atomistic origin for $\mu=2$.

\subsection{Intrinsic KAI coercive fields from multiscale simulations}

\begin{figure}
	\begin{center}
		\includegraphics[width=0.9\textwidth]{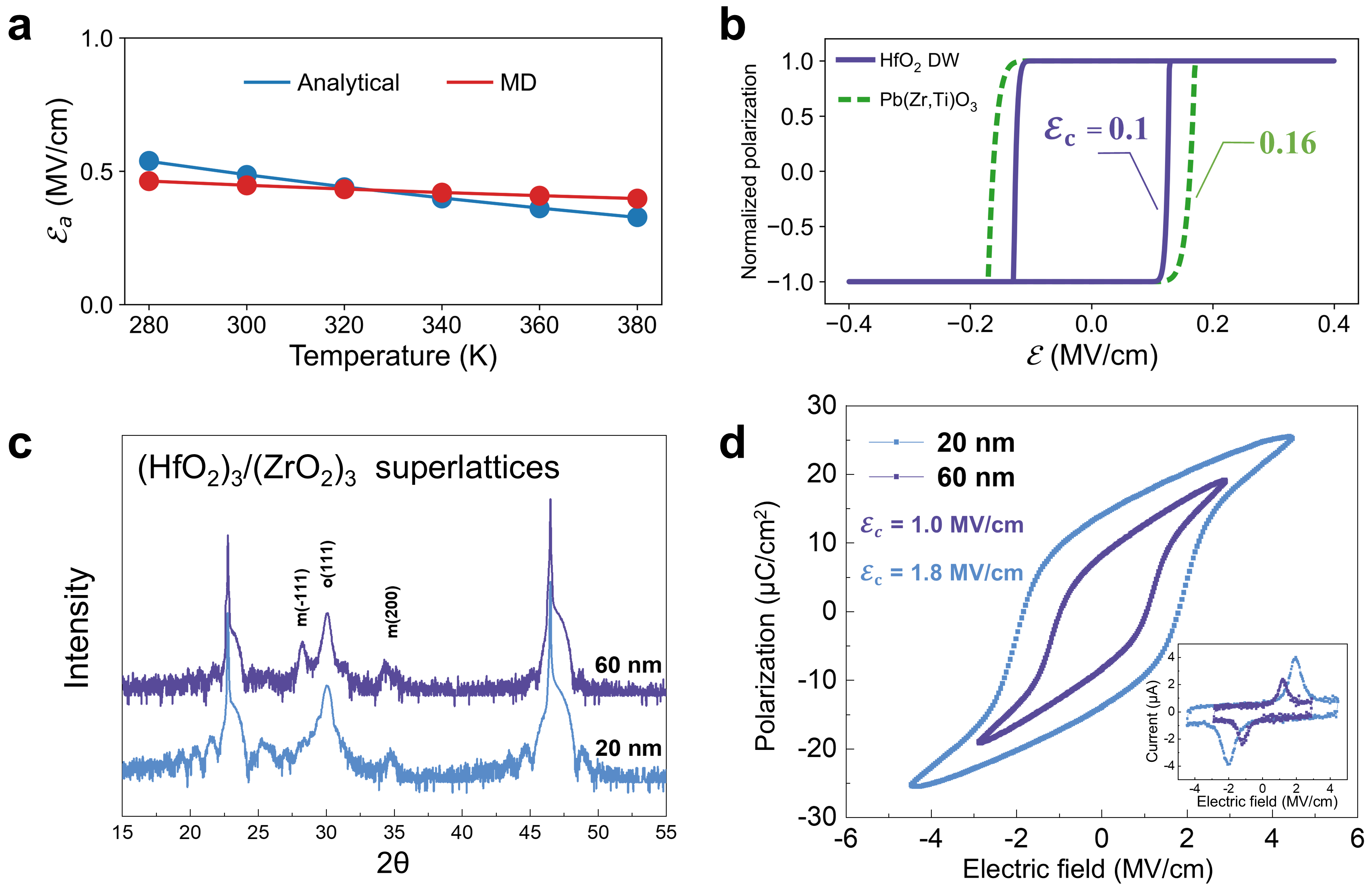}
	\end{center}
	\caption{\textbf{KAI coercive fields from multiscale simulations and experiments.} \textbf{a,} Activation fields ($\mathcal{E}_a$) obtained from MD simulations compared with predictions from the LGD analytical model. \textbf{b,} Simulated intrinsic polarization-electric field hysteresis loops in HfO$_2$ thin films, driven by the motion of $Pbcn$-type walls. For comparison, the simulated hysteresis loop for Pb(Zr,Ti)O$_3$ thin films is also included. \textbf{c,} XRD $\theta$-2$\theta$ patterns for (HfO$_2$)$_3$/(ZrO$_2$)$_3$ superlattice heterostructures with total film thickness of 20 nm and 60 nm. \textbf{d,} Polarization-electric field hysteresis loops for the superlattice heterostructures with insets showing the switching current-electric field curves.
 } 
 \label{KAIEc}
\end{figure}

As the domain wall motion governs the KAI-type switching, the polarization-electric field ($\mathcal{P}$-$\mathcal{E}$) hysteresis loop that determines the value of \Ec~can be simulated based on the field-dependent domain wall velocity~\cite{Liu16p360}. A key parameter needed in Equation~(\ref{merz}) is the activation field $\mathcal{E}_a$, which is directly linked to the critical nucleation barrier (\Unuct) from the LGD 2D nucleation model. 
Using Avrami's theory of transformation kinetics, we derive the relationship between \Unuct~and $\mathcal{E}_a$ for domain wall motions with $\mu=2$ as 
$\mathcal{E}_a=\sqrt{\frac{1}{D+1}\frac{\lambda\Delta U_{\rm nuc}^*}{k_BT}}\mathcal{E}$, where $D$ represents the dimensionality and $\lambda$ is a scaling parameter mapping $\Delta U_{\rm nuc}^*$, associated with a half-unit-cell hop, to the effective barrier encountered by a domain wall moving one unit cell at a time (see Supplementary Sect.~V). For $D=2$ and $\lambda=0.5$, we calculate $\mathcal{E}_a$ values using the LGD model at various temperatures, and the analytical predictions agree well with MD results (Fig.~\ref{KAIEc}a). 
Leveraging the LGD model's efficient prediction of $\mathcal{E}_a$, we can accurately simulate hysteresis loops driven by domain wall motion to determine \Ec. As depicted in Fig.~\ref{KAIEc}b, the $\mathcal{P}$-$\mathcal{E}$ loop for $Pbcn$-type walls in HfO$_2$ yields $\mathcal{E}_c \approx$0.1 MV/cm, even lower than the $\approx$0.2 MV/cm due to 180$^\circ$ domain walls in Pb(Zr,Ti)O$_3$ thin films~\cite{Liu16p360}. 

The theoretical lower limit of 0.1 MV/cm for KAI-type switching underscores the potential of domain-wall engineering to significantly reduce \Ec. One approach to activate KAI-type switching is to increase the film thickness. However, as shown in Fig.~\ref{Thinfilm}, thicker films (\textit{e.g.}, 20 nm) exhibit a substantial volume fraction of the nonpolar monoclinic phase.
To address this challenge, we employ artificially-designed (HfO$_2$)$_3$/(ZrO$_2$)$_3$ superlattice heterostructures, which enable the realization of thicker ferroelectric films while suppressing the formation of the monoclinic phase.
As shown in Fig.~\ref{KAIEc}c, both 20 nm and 60 nm superlattices display a strong $(111)_o$ peak near 30$\degree$, corresponding to the polar $o$-phase, alongside minor $m$-phase peaks at 28.2$\degree$ and 34.5$\degree$. The 20 nm superlattice demonstrates high $o$-phase stability with minimal $m$-phase content, reflecting excellent growth quality. Although the $m$-phase fraction increases in the 60 nm film, the $o$-phase remains dominant. This is confirmed by $\mathcal{P}$-$\mathcal{E}$ loops and switching current-electric field curves (Fig.~\ref{KAIEc}d), showing clear ferroelectric hysteresis and switching peaks. Notably, the 60 nm superlattice achieves a remnant polarization $>8$ $\mu$C/cm$^2$ with a low \Ec~of 1~MV/cm, much lower than the 1.8~MV/cm in the 20 nm superlattice. This supports geometry optimization that likely promotes domain wall motions as potential strategy to activate KAI-type switching and reduce \Ec.

\section{Conclusions}
The multiscale approach developed in this work integrates DFT calculations, MD simulations, and the analytical LGD model, enabling precise predictions of coercive fields \Ec~in ferroelectric hafnia for both NLS- and KAI-type switching. Atomistic insights into the critical nucleus, particularly its shape and boundary lattice structure identified through MD simulations, are crucial for the accuracy of the LGD model. The theoretical \Ec~values for NLS, predicted using multiscale simulations, successfully reproduce the scaling of coercive fields for film thicknesses ranging from 3 to 20 nm. We propose the sample-dependent reverse size effect, manifested as thickness-induced strain, explains the diverse scaling relationships reported in the literature.

Our combined theoretical and experimental investigations reveal that the giant \Ec~in hafnia-based ferroelectrics arises from the ultra-thin geometry, which confines switching to the NLS mechanism. In contrast, the theoretical lower limit for KAI-type \Ec~is remarkably low, at 0.1 MV/cm, rivaling that of perovskite ferroelectrics. This low value is associated with $Pbcn$-type 180$\degree$ domain walls, whose motion involves a two-hop mechanism and the formation of a half-unit-cell-thin elliptical nucleus. The unconventional creep exponent of $\mu=2$ for domain wall dynamics also arises from this mechanism. A promising strategy to enable KAI-type switching is to increase the film thickness to promote domain wall motion. This approach is validated by our experimental demonstration of a low coercive field of 1 MV/cm in a 60 nm PLD-grown ferroelectric (HfO$_2$)$_3$/(ZrO$_2$)$_3$ superlattices.
These findings provide a deeper understanding of the mechanisms governing ferroelectric switching in hafnia-based ferroelectrics and highlight the potential for achieving low coercive fields through geometry optimization and domain-wall engineering.

\clearpage
\newpage 
\section {Methods}

\subsection {DFT calculations} 

All first-principles density functional theory (DFT) calculations are performed using the Vienna \textit{ab initio} simulation package (\texttt{VASP})~\cite{Kresse96p11169, Kresse96p15} employing the Perdew-Burke-Ernzerhof (PBE) density functional~\cite{Perdew96p3865}. An energy cutoff of 600 eV and a $k$-spacing of 0.5 \AA$^{-1}$ are sufficient to converge energy within $10^{-6}$ eV and atomic forces within $10^{-2}$ eV/\AA. The optimized lattice parameters for the unit cell of $Pca2_1$ HfO$_2$ are $a=5.266$ \AA, $b=5.047$ \AA, and $c=5.077$ \AA, with polarization directed along the $c$ axis ($z$ axis). We identify the minimum energy path (MEP) for polarization switching at the unit cell level using the DFT-based variable-cell nudged elastic band (VCNEB) technique and a 12-atom unit cell consisting of four hafnium and eight oxygen atoms.
During this process, lattice constants are allowed to relax.
The domain wall energy ($\sigma$) at 0~K is estimated using the following equation:
\begin{equation}
    \sigma = \frac{E_{\rm DW}-E_{\rm bulk}}{2S_{\rm DW}},
\label{sigma}
\end{equation}
where $E_{\rm DW}$ is the total energy of a $1\times24\times1$ supercell containing the domain walls (see Fig.~S1a), $E_{\rm bulk}$ is the energy of an equivalent-sized single-domain $Pca2_1$ HfO$_2$ supercell, and $S_{\rm DW}$ denotes the domain wall area. The atomic positions and lattice constants of the supercell are fully optimized.
Because of periodic boundary conditions employed in DFT calculations, each supercell includes two DWs, which is accounted for by the factor of 2 in the denominator of Equation~(\ref{sigma}). The DFT values of domain wall energies are reported in Table.~S1.

\subsection{Deep potential from DFT} 

The deep neural network-based model potential, referred to as deep potential (DP), employed in MD simulations maps an atom's local environment to its energy ($E_i$), the sum of which gives the total energy $E$, $E=\sum_iE_i$. 
The training database includes energies and atomic forces of 21,790 HfO$_2$ configurations and 34,222 oxygen-deficient HfO$_{2-\delta}$ configurations,  96 domain wall configurations, all computed with DFT. These configurations span various phases, including $P2_1/c$, $Pbca$, $Pca2_1$, and $P4_2/nmc$, as well as some intermediate states during polarization switching structures. Figure~S3 presents a comparison between the energies and atomic forces for all configurations in the training database as calculated by both DP and DFT, demonstrating a high level of agreement between DFT and DP calculations. The DP model exhibits a mean absolute error of 3.17 meV/atom for energy and 0.15 eV/
\AA~for atomic force. Additional tests confirming the accuracy of the DP model are presented in Supplementary Sect.~II. 
In an effort to improve reproducibility, we have developed an \note{online {\href{https://nb.bohrium.dp.tech/user/update/52795761357}{notebook}}}~\cite{notebookforL35}
to share the training database, force field model, and training metadata. 

\subsection{MD simulations} 

To establish the electric field-dependence of 180$^\circ$ domain wall velocity ($v$) in ferroelectric HfO$_2$, we perform isobaric-isothermal ($NPT$) ensemble MD simulations with a deep potential of hafnia over a wide range of temperatures and electric fields using \texttt{LAMMPS}~\cite{Plimpton95p1}. 
Temperature control is achieved using the Nos\'e-Hoover thermostat, while pressure is maintained through the Parrinello-Rahman barostat. The time step for the simulations is set at 2 fs. To account for the effects of electric fields in classical MD simulations, we adopt the force method~\cite{Umari02p157602}. This approach involves adding an extra force to each ion, calculated by multiplying the ion's Born effective charge by the electric field's intensity. 
We use a $8\times24\times8$ supercell comprising 18432 atoms to model 180$^\circ$ domain walls, which separate two domains with opposite polarization along the $z$ axis. The supercell is first equilibrated at the designated temperature for a minimum duration of 50 ps. Following this, the electric field along the $z$ axis is activated.
The value of $v$ is determined directly by tracking the distance traveled by the wall within a specific time frame.
Considering the stochastic nature of nucleation, we performed at least 20 simulations with slightly varying initial configurations at each set temperature and electric field to determine average velocities.

\subsection{Determine critical nucleus with MD-based $r$PEM }  

Conventional MD simulations are limited in capturing nucleation at the domain wall under low-strength electric fields, as such events are too rare over the accessible time scales. To address this, we have adapted the persistent-embryo method (PEM), originally developed for modeling 3D crystal nucleation from liquids~\cite{Sun18p085703}. Our modification, referred to as the revised PEM ($r$PEM), is specifically tailored for simulating electric field-driven nucleation in ferroelectrics.
In the $r$PEM approach, a 3D embryo containing $N_0$ switched oxygen atoms is initially embedded within a single-domain configuration. Similarly, for simulating nucleation at a wall, a 2D embryo comprising $N_0$ switched oxygen atoms is embedded within the unswitched domain wall region.
During the simulation, if the position of an oxygen atom changes by more than 0.3~\AA~relative to its initial position, it is then classified as switched and included in the embryo. 
A tunable harmonic potential is applied to each oxygen atom of the embryo to prevent back-switching. 
This strategy effectively suppresses long-term nucleation fluctuations in smaller nuclei. The potential is progressively weakened and ultimately removed once the nucleus size ($N$) exceeds a subcritical threshold ($N_{\rm sc}$). The spring constant of the harmonic potential follows the formula $k(N) = k_0\left[(N_{\rm sc} - N)/N \right] $ if $N<N_{\rm sc}$, and becomes zero $k(N)=0$ otherwise, as sketched in Fig.~S2.
By using an adaptive spring constant that reduces to zero before the critical nucleus size is reached, \textit{r}PEM ensures that the system's dynamics are unbiased at the critical point.
The observed plateau on the nucleus size versus time curve, $N(t)$, then indicates the period during which the nucleus size oscillates around $N^*$. This fluctuation allows us to determine the shape and size of the critical nucleus in an unbiased environment. 

\subsection{Pulsed-laser deposition of thin-film heterostructures}
Heterostructures consisting of 10 nm LaSr$_{0.67}$Mn$_{0.33}$O$_3$ (LSMO)/$x$ nm Hf$_{0.5}$Zr$_{0.5}$O$_2$ (HZO) films ($x$ = 5–12 nm) and 10 nm LSMO/$y$ nm (HfO$_2$)$_3$/(ZrO$_2$)$_3$ superlattices ($y$ = 20 and 60 nm) were epitaxially grown on (001)-oriented single-crystal STO substrates using pulsed laser deposition (PLD, Arrayed Materials RP-B). The (HfO$_2$)$_3$/(ZrO$_2$)$_3$ superlattices were fabricated by alternating deposition from HfO$_2$ and ZrO$_2$ targets.
The bottom LSMO electrode was deposited at a substrate temperature of 700°C with a laser fluence of 0.8 J/cm$^2$ and a repetition rate of 3 Hz. The ferroelectric HZO films and (HfO$_2$)$_3$/(ZrO$_2$)$_3$ superlattices were subsequently deposited at 600°C using a laser fluence of 1.3 J/cm$^2$ and a repetition rate of 2 Hz for all targets. After deposition, the heterostructures were cooled to room temperature at 10°C/min under a static oxygen pressure of 10$^4$ Pa. Circular platinum (Pt) top electrodes, 90 nm thick, were deposited via magnetron sputtering (Arrayed Materials RS-M) after photolithography, with varying areas to suit different measurements.
\begin{acknowledgments}
J.Y., J.W., and S.L. acknowledge the supports from National Key R\&D Program of China (2021YFA1202100), National Natural Science Foundation of China (12074319, 12361141821,), and Westlake Education Foundation. 
J.L., C.Z., and Z.C. acknowledge the support from National Natural Science Foundation of China (92477129).
The computational resource is provided by Westlake HPC Center.
\end{acknowledgments}
\clearpage
\newpage




\newpage
\bibliography{SL.bib}

\clearpage
\newpage

\clearpage
\newpage

\clearpage
\newpage

\clearpage
\newpage

\clearpage
\newpage

\clearpage
\newpage

\clearpage
\newpage

\end{document}